\documentclass[]{emulateapj}
\usepackage{natbib}
 
\def\snep{SNe\,II-P} 
 
\def\fe{\ion{Fe}{2}}
\def\hb{{H}{${\beta}$}}
\def\s2n{S/N}
\def\mqb{Malmquist bias}

\begin{document}

\submitted{ApJ Accepted \today} 

\title {Type II-P Supernovae as Standard Candles: The SDSS-II Sample Revisited}
\shorttitle{SDSS II SNe\,II-P revisited}
\shortauthors{Poznanski, Nugent, \& Filippenko}

\author{Dovi Poznanski\altaffilmark{1,2,3}, 
Peter E. Nugent\altaffilmark{1}, and
Alexei V. Filippenko\altaffilmark{2}}

\email{dovi@berkeley.edu}

\altaffiltext{1}{Lawrence Berkeley National Laboratory, 1 Cyclotron
  Road, Berkeley, CA 94720.}
\altaffiltext{2}{Department of Astronomy, University of California,
  Berkeley, CA 94720-3411.}
\altaffiltext{3}{Einstein Fellow.}

\begin{abstract}

We revisit the observed correlation between \hb\ and \fe\ velocities
for Type II-P supernovae (SNe~II-P) using 28 optical spectra of 13
SNe~II-P and demonstrate that it is well modeled by a linear relation
with a dispersion of about 300 km\,s$^{-1}$.  Using this correlation,
we reanalyze the publicly available sample of SNe~II-P compiled by
\citeauthor{dandrea10} and find a Hubble diagram with an intrinsic
scatter of 11\% in distance, which is nearly as tight as that measured
before their sample is added to the existing set. The larger scatter
reported in their work is found to be systematic, and most of it can
be alleviated by measuring \hb\ rather than \fe\ velocities, due to
the low signal-to-noise ratios and early epochs at which many of the
optical spectra were obtained.  Their sample, while supporting the
mounting evidence that SNe~II-P are good cosmic rulers, is biased
toward intrinsically brighter objects and is not a suitable set to
improve upon SN~II-P correlation parameters. This will await a
dedicated survey.

\end{abstract}


\defcitealias{dandrea10}{D10}
\defcitealias{nugent06}{N06}
\defcitealias{poznanski09}{P09}

\section{Introduction}

Type II supernovae (SNe) that undergo a long, bright, flat phase in
photometric evolution are commonly referred to as plateau SNe
(\snep). These have been shown over the past decade to be good
``standardizable candles,'' and potential cosmological probes, in a
manner similar to that of their more famous cousins,
SNe~Ia. \citet{hamuy02} were the first to demonstrate this empirical
``standardizable candle method'' (SCM), a technique which was later
streamlined and applied to larger samples by \citet[][hereafter
  N06]{nugent06} and \citet[][hereafter P09]{poznanski09a}; see also
\citet{olivares10}.

P09 used a sample of 34 \snep\ to constrain the three parameters that
define the correlation: a reference absolute magnitude in the $I$
band, a velocity term that is correlated with the luminosity during
the plateau, and a color term that minimizes the contribution from
intrinsic color inhomogeneity as well as from dust extinction. The
resulting intrinsic scatter in the Hubble diagram was found to be
0.22\,mag, which is equivalent to about 10\% in distance, similar to
the result of N06 despite the larger sample. Since the samples used
by N06 and P09 are dominated by nearby SNe that sport large distance
uncertainties due to peculiar velocities, a robust derivation of the
SCM parameters, together with a determination of its potential power,
requires a set of SNe~II-P in the Hubble flow.

Recently, \citet[][hereafter D10]{dandrea10} presented such a sample
of 15 \snep\ in the Hubble flow from the Sloan Digital Sky Survey II
(SDSS-II) SN survey, albeit obtained during a project that was focused
on SNe~Ia. The SCM method relies on photospheric velocities measured
from the \fe\ $\lambda$5169 absorption line during the plateau phase,
specifically about 50 days past explosion. As noted by D10, however,
the optical spectra of their SNe~II-P are usually of low
signal-to-noise ratio (\s2n) and taken at an early epoch, when
\fe\ lines are not fully developed. This observation strategy, while
complicating the application of the SCM method to \snep, was dictated
by the main purpose of their survey, which was to identify SNe~Ia and
measure their redshifts in order to use them as distance indicators
\citep{kessler09}. D10 find that adding their sample to those 
analyzed previously increases the scatter significantly, to 15\% in
distance. This points to a failure of the data, method, or underlying
model.

While D10 conscientiously explore many possible sources for this
scatter, we show below that most of it is due to a systematic offset,
and could be explained by two effects: the method used to determine
the photospheric velocities, and an observational bias. In
\S\ref{s:vel} we rederive photospheric velocities for their sample
using \hb\ absorption lines, as applied by N06 to their sample of
higher redshift SNe. We calibrate the relation between \hb\ velocities
and the standard \fe\ velocities in \S\ref{s:hb}. In \S\ref{s:hd} we
update the Hubble diagram and discuss the specific selection effects
that bias the SDSS-II sample.

\begin{figure}[!ht] 
\includegraphics[width=3.0in]{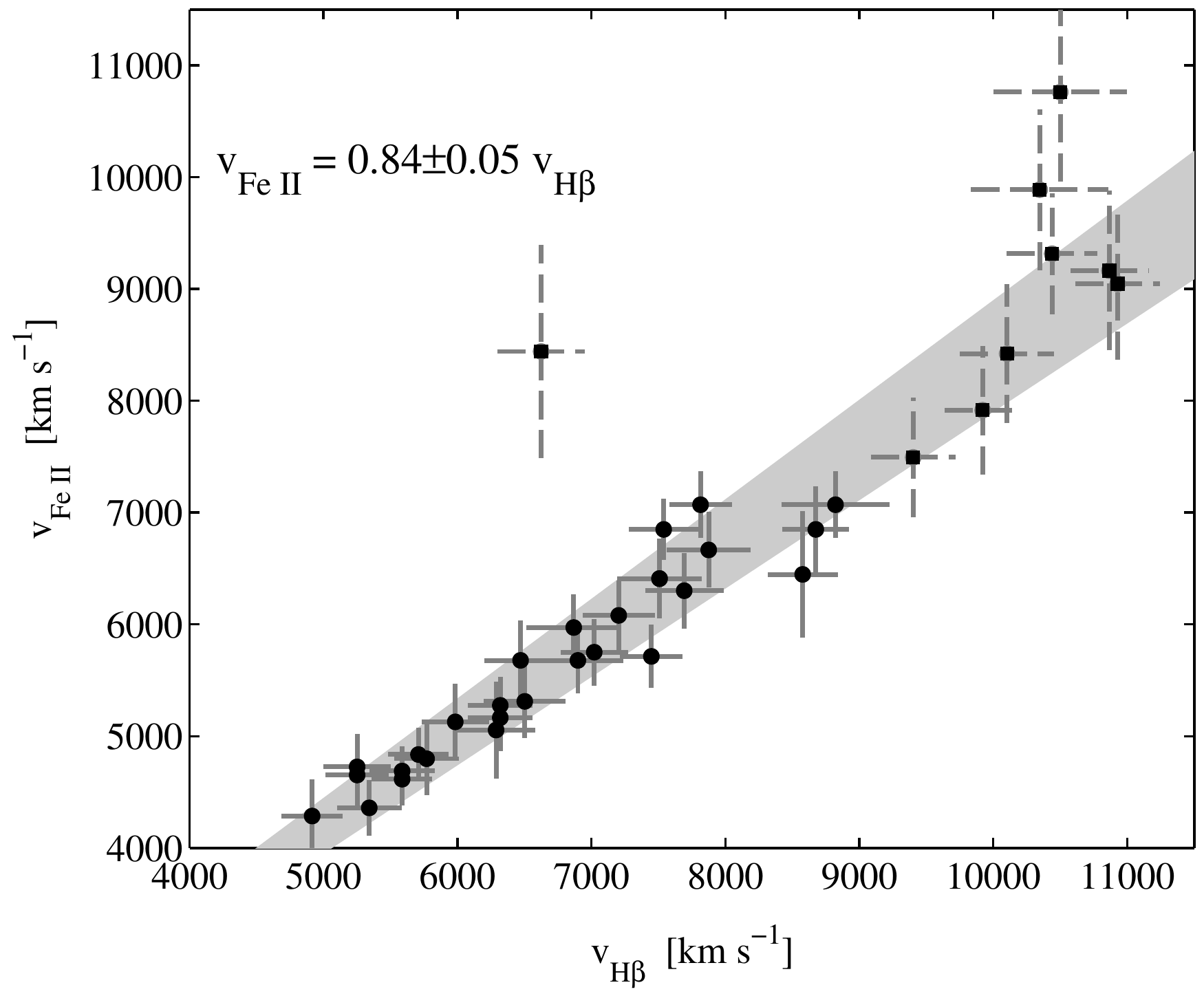}
\caption{Velocities determined from the absorption minima of
  \hb\ $\lambda$4861 and \fe\ $\lambda$5169 using 28 spectra of 13
  \snep\ at phases of 5 to 40 days after explosion. The shaded area
  marks the 1$\sigma$ region for the derived correlation, $v_{\rm
    Fe~II} = (0.84 \pm 0.05)\,v_{\rm H\beta} $. \fe\ velocities can be
  well determined using \hb, which is easier to detect in early-phase
  and low-\s2n spectra. Squares with dashed error bars mark
  \fe\ velocities derived for objects with very high \hb\ velocities
  using Eq. 2 of N06. The outlier at $v_{\rm Fe~II} \approx
  8,400$\,km\,s$^{-1}$ is from a very early spectrum (2 days past
  explosion) of the low-luminosity, low-velocity
  SN\,2005cs.}\label{f:hbeta}
\end{figure}

\section{Photospheric Velocities}\label{s:vel}

In order to derive photospheric velocities, D10 apply to their spectra
the algorithm developed by P09. Briefly, using the ``SN Identification
Code'' (SNID; \citealt{blondin07}), the spectra are cross-correlated
with a library of high-\s2n\ templates for which \fe\ velocities can
be measured precisely. P09 successfully applied this method to 19
nearby SNe~II-P that typically had a few high-\s2n\ spectra per
object, many of them near day 50 after explosion. D10 correctly point
out that this method may not be applicable to their sample, due to the
early epoch at which they were obtained and their low \s2n. The
\fe\ lines usually become fully developed after a few weeks past
explosion. Even at later times those lines are often weak, making
velocity derivation from low-\s2n\ spectra problematic.

We rederive the velocities for the SDSS-II sample using another
method, and find values that are typically different from those of
D10. These in turn improve the scatter in the Hubble diagram, as
demonstrated in \S\ref{s:hd}. N06 showed that there is a correlation
between the velocities measured from \hb\ $\lambda$4861 and those
derived from \fe\ $\lambda$5169. In the next section we improve the
determination of that correlation using many more objects and spectra.

We obtain the photospheric velocities for the SDSS-II sample as
follows. From each spectrum we measure the \hb\ velocity, and where
feasible, the \fe\ velocity. This is done by finding the minimum of
the absorption line. When the \s2n\ is too low we smooth the spectra
with a Savitzky-Golay filter (which is similar to a running mean;
\citealt{savitzky64}). For most of the SDSS-II spectra,
\fe\ $\lambda$5169 is either undeveloped or buried in the noise. The
\hb\ velocities are translated to equivalent \fe\ velocities using the
relation found in \S\ref{s:hb} (with the uncertainty in that
relationship folded in quadrature). For every spectrum that does have
a direct \fe\ velocity measurement, we calculate a weighted mean
between the direct and indirect values. These velocities and
uncertainties are then propagated to day 50 (as was done by N06, P09,
and D10) and listed in Table \ref{t:vel}.

\begin{deluxetable}{lcc}
\tablewidth{0pt}
\tabletypesize{\scriptsize}
\tablecaption{Day-50 Velocities for the SDSS-II SNe II-P\label{t:vel}}
\tablehead{
\colhead{IAU Name} &
\colhead{D10 (km\,s$^{-1}$)} &
\colhead{\hb\ based (km\,s$^{-1}$)} }
\startdata
SN 2007ld & 4110(570)  &  3970(660) \\
SN 2006iw & 4490(160)  &  4660(270) \\
SN 2007lb & 3950(340)  &  4080(420) \\
SN 2007lj & 4220(460)  &  4110(450) \\
SN 2006jl & 5560(310)  &  4780(350) \\
SN 2007lx & 3900(820)  &  4170(910) \\
SN 2007nw & 4290(750)  &  4830(840) \\
SN 2006kv & 3640(620)  &  4050(950) \\
SN 2007kw & 3930(460)  &  4990(1060) \\
SN 2006gq & 3400(150)  &  3790(440) \\
SN 2007ky & 3510(420)  &  4060(480) \\
SN 18321\tablenotemark{a}  & 5060(640)  &  5820(700) \\
SN 2006kn & 4220(440)  &  5000(590) \\
SN 2007kz & 4380(320)  &  5490(370) \\
SN 2007nr & 4430(350)  &  4900(590) \\

\enddata 
\tablecomments{Values in parentheses are 1$\sigma$ uncertainties.}
\tablenotetext{a}{Internal SDSS name from D10.}
\end{deluxetable}

\section{Correlation between \hb\ and \fe\ Velocities}\label{s:hb}

N06 propose an alternative velocity proxy to \fe, using the
\hb\ absorption line that is both present early and significantly more
prominent. We repeat the analysis of N06 using the much larger sample
of P09 and derive a new relation between the two velocities.

We use 28 spectra of 13 \snep\ that span ages of 5 to 40 days past
explosion; both lines are readily identifiable in all of the
spectra. We restrict ourselves to spectra obtained prior to day 40,
since at later times \fe\ $\lambda$5169 is at least as strong as \hb,
which also becomes blended with other lines that can bias the velocity
measurement. We measure the velocities as described for the SDSS-II
SNe in \S\ref{s:vel}.

As seen in Figure~\ref{f:hbeta}, we find a robust ($\chi^2/{\rm dof}
\approx 1$) linear correlation that can be represented as $v_{\rm
  Fe~II} = (0.84 \pm 0.05)\,v_{\rm H\beta} $. The uncertainty is
equivalent to an additional error of about 200--400\,km\,s$^{-1}$ over
the relevant range of velocities. While this relation is somewhat
different from the one derived by N06, in practice it gives similar
results for the range of velocities probed by the SDSS-II sample.

Since there are very few objects in the P09 sample having velocities
higher than 8,000\,km\,s$^{-1}$ in which both lines are well
developed, we add to Figure~\ref{f:hbeta} nine early-epoch,
high-velocity spectra where the \fe\ lines are not detected. In order
to approximate the \fe\ velocity at those times, we use the velocity
derived for these objects at day 50 in P09, and propagate them to the
correct epoch using Equation 2 of N06 which models the time evolution
of \fe\ velocities. Most of the additional spectra seem to agree with
the relationship derived above, with two notable exceptions. One is
SN\,2005cs (2 days past explosion) with a derived \fe\ velocity which
is much too high. This object was a low-luminosity, low-velocity
SN~II-P, and is of little relevance to nonlocal samples. The second
exception is that at \hb\ velocities higher than 10,000\,km\,s$^{-1}$
the scatter increases significantly. This may be due to a combination
of line blending at these high velocities, combined with very blue
continua that make absorption minima harder to define and measure. All
of the SNe in the D10 sample have a spectrum with \hb\ velocity
smaller than 10,000\,km\,s$^{-1}$.  While these exceptions are not
relevant to the reanalysis of the D10 sample, one should be cautious
and remember that this correlation may break down for spectra with
extremely high or low velocities.

\begin{figure*}[t] 
	\center
\includegraphics[width=6.in]{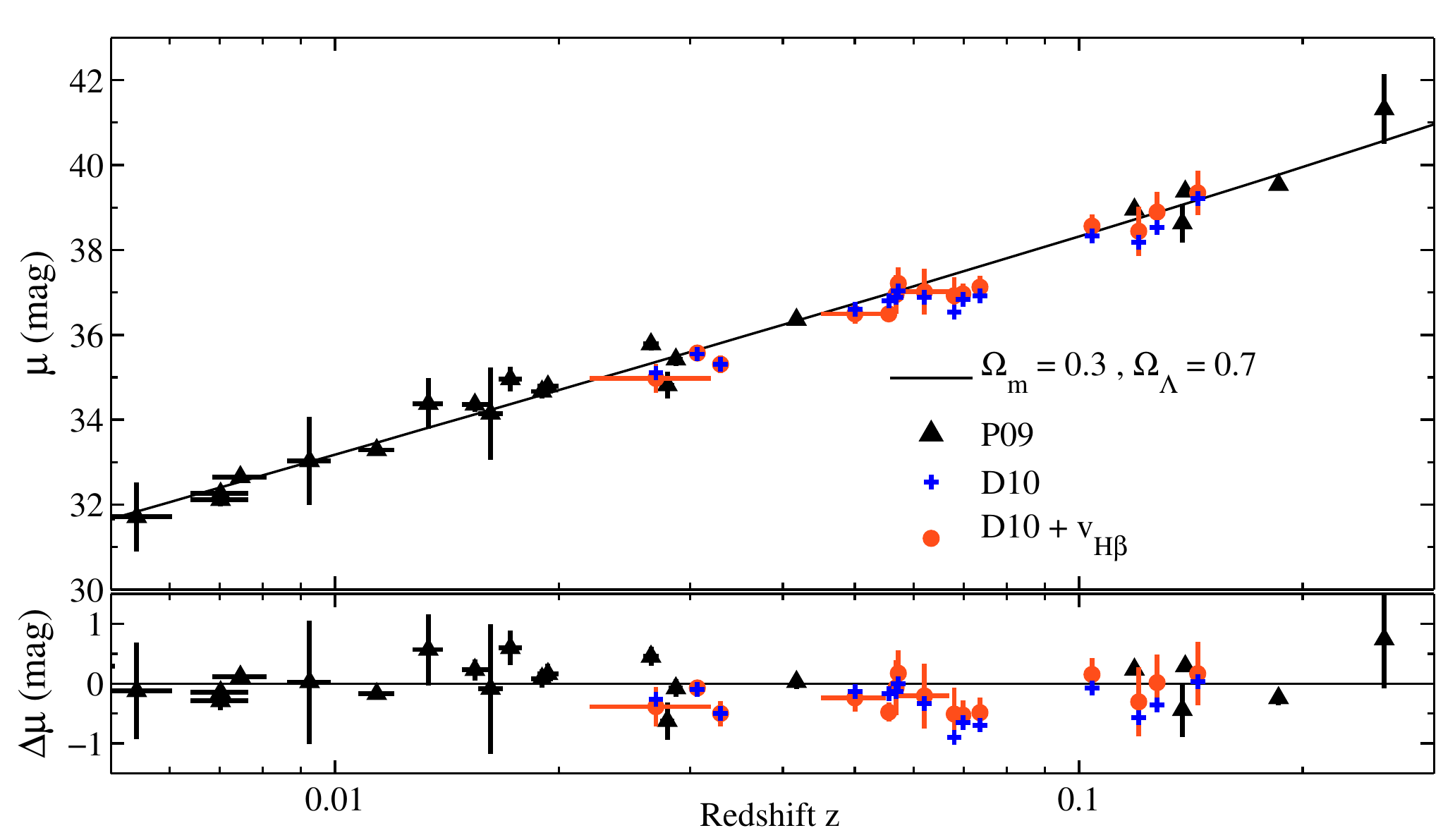}
\caption{Hubble diagram (top) and residuals from the standard
  concordance cosmological model (bottom) for the P09 sample (black
  triangles), the D10 sample (blue crosses; uncertainties omitted for
  clarity), and the D10 sample with corrected velocities (orange
  filled circles). The standard cosmological model (line) is shown to
  guide the eye. The velocity correction mostly shifts the D10 sample
  to fainter luminosities, and hence to larger ``apparent distances,''
  more consistent with the Hubble law. For clarity we show here only
  the sample of 36 SNe~II-P at redshifts $z > 0.005$, where
  peculiar-velocity uncertainties are less
  dominant.} 
\label{f:hd} \vspace{.2cm}
\end{figure*}

\section{Hubble Diagram}\label{s:hd}

When minimizing the ``cost function'' (Equation 4 of P09) in order to
determine the best-fit parameters, P09 find an intrinsic scatter of
0.22 mag (D10 measure it to be 0.20 mag when minimizing a slightly
different function), which is equivalent to $\sim$10\% in distance,
similar to the result of N06. D10 report an increase to 0.29 mag
(which we measure to be 0.32 mag) when adding the SDSS-II sample,
which is about 50\% more scatter. An important clue to the source of
this additional dispersion is provided by Figure 5 of D10, which
clearly shows that the SDSS-II SNe are systematically offset from the
Hubble law, with all but one of them falling below the line. In fact,
the scatter is reduced significantly by artificially dimming the
SDSS-II SNe by 0.4 mag, ending up even smaller than that reported by
P09.

We repeat the analysis of D10 using all of their data, except for the
velocities and their respective uncertainties, where we substitute
those we find in \S\ref{s:vel}. As can be seen in Figure~\ref{f:hd},
this eliminates a substantial fraction of the systematic offset, which
in turn reduces the intrinsic scatter to 0.25 mag (11\% in distance),
only slightly higher than without this sample. Our best-fit parameters
for the combined sample are nearly identical to those of P09, except
for the reference absolute magnitude which is 0.1 mag brighter for the
combined sample.

Deriving the parameters using only the SDSS-II sample, we still find
preferred values for the correlation coefficients that are offset from
those of P09. The SDSS-II SNe appear to be predominantly overluminous,
and favor a small value for the velocity correlation coefficient. 
Following the suggestion made by D10, we show below that this is 
largely due to an observational bias arising from the SDSS-II 
follow-up criteria.

We have performed Monte Carlo simulations, starting from a SN
population that mimics the properties of the P09 sample, to which we
apply various observational cuts. For these simulated samples we then
compute the best-fit correlation parameters. We find that a simple
magnitude cut cannot produce such biased values. However, selecting
objects that are \emph{intrinsically brighter} skews the parameter
derivation in the right direction. Folding in the larger velocity
uncertainties allows us to reproduce the bias toward luminous SNe with
little correlation. Indeed, when one selects only luminous SNe, it is
not surprising that the best-fit reference absolute magnitude is
brighter, and the correlation parameters smaller --- thus pointing
toward a weaker link between luminosity and velocity. The lack of
statistical power of such a sample is reflected by the fact that the
parameters derived from the combined set are almost identical to those
recovered without including the SDSS-II sample. Because the SDSS-II
sample does not cover a substantial region of parameter space, it
cannot significantly constrain the SCM parameters.

\begin{figure}[!ht] 
\center
\includegraphics[width=3.0in]{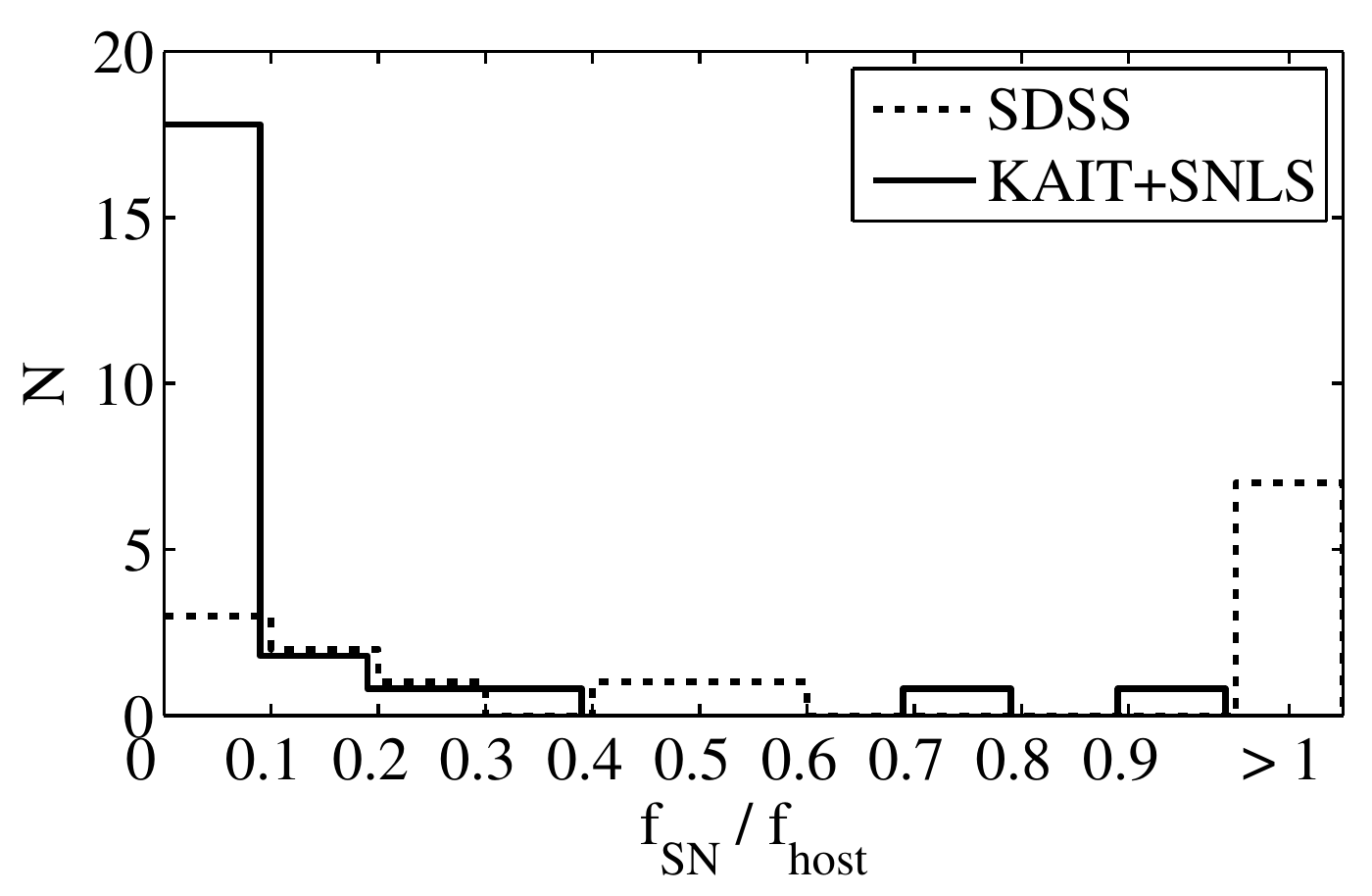} \caption{Approximate $I$-band
  flux ratios (SN/host galaxy) for the Katzman Automatic Imaging
  Telescope (KAIT; P09) and Supernova Legacy Survey (SNLS; N06)
  samples (solid line) and for the SDSS-II sample from D10 (dotted
  line). About half of the SDSS-II SNe are brighter than their hosts
  galaxies, a selection effect that removes intrinsically faint
  objects and skews the estimation of correlation
  parameters.} \label{f:contrast}
\end{figure}

D10 mention that the contrast between a SN and its host galaxy was a
selection criterion for spectroscopic follow-up observations. This 
was obviously done to increase the efficiency of their spectroscopic 
observing runs and the number of SNe~Ia they find. Unlike a
regular \mqb\ that skews the distribution near the detection limit of
the survey, such a criterion selects intrinsically bright objects,
independent of redshift. This is evident from the very different
luminosity functions of the two samples (Fig. 7 of
D10). Figure~\ref{f:contrast} shows that nearly half of the SDSS-II
SNe are brighter than their host galaxies up to a factor of a few,
unlike the samples of P09 and N06 that are consistently fainter than
their hosts. A better understanding of the SCM relation, based on
larger, less biased samples, will allow a careful determination of the
various potential biases, providing a crucial component for
constraining cosmological parameters.

\section{Conclusions}\label{s:conc}

We find that when measuring expansion velocities for the SDSS-II
sample of \snep\ by using \hb, the SCM correlation gives a tight
Hubble diagram, with a scatter of $\sim$11\% in distance. This is
comparable to the scatter obtained with previous samples of \snep,
despite a systematic offset among the SDSS-II objects, which is
probably due to a strong bias favoring intrinsically bright SNe.

Our analysis shows that \hb\ velocities correlate well with those
derived from \fe\ lines, with a scatter of about 300\,km\,s$^{-1}$,
enabling the use of early-time spectra and data having low \s2n.

The SCM parameters and the true tightness of the correlation remain to
be tested with high-quality Hubble-flow data. We are currently
compiling such a set of \snep\ from the Palomar Transient Factory
(\citealt{rau09,law09,poznanski09b}), where cosmology with \snep\ is
one of the key projects and should lead to less biased samples.

\acknowledgments 

We thank C. D'Andrea and his collaborators for making the SDSS-II data
on SNe~II-P available to the community, and J. S. Bloom for useful
advice. D.P. is supported by an Einstein Fellowship. We acknowledge
support from the US Department of Energy Scientific Discovery through
Advanced Computing (SciDAC) program under contract
DE-FG02-06ER06-04. A.V.F. is also grateful for funding from National
Science Foundation grant AST-0908886 and the TABASGO Foundation.

\end{document}